\documentstyle[12pt]{article}
\begin {document}

\begin{center}

     {\bf  Quantum Hamilton-Jacobi formalism \\ and the bound state spectra }

     \vspace{1 cm}R.S. Bhalla, 
     A.K. Kapoor \footnote{$^{}$ {email: akksp@uohyd.ernet.in}}
and      P.K. Panigrahi \footnote{$^{}${email: panisp@uohyd.ernet.in}}
     \vspace{1.5 cm}

   School of Physics,\\ University of Hyderabad,\\ Hyderabad-500 046, INDIA.

\vspace{2 cm}

{\bf Abstract}
\vspace{0.5 cm}
\end{center}


It is well known in classical mechanics that, the frequencies of a periodic 
system  can be obtained rather easily through the action variable, without 
completely  solving the equation of motion. The equivalent quantum action 
variable appearing in the quantum Hamilton-Jacobi formalism, can, analogously 
provide  the  energy eigenvalues of a bound state problem, without having to 
solve  the corresponding Schr\"odinger equation explicitly. This elegant and 
useful method is elucidated here in the context of some known and not so 
well known solvable potentials. It is also shown, how this method provides an  
understanding, as to why approximate quantization schemes such as  ordinary  
and  supersymmetric WKB, can give exact answers for certain potentials. \\

\vspace { 5.5 cm}
\noindent
December 1995
\newpage 
\pagestyle{plain}
\baselineskip = 24pt
\noindent
{\bf I. INTRODUCTION } \\
\smallskip 
In classical mechanics, the Hamilton-Jacobi (H-J) theory is a
well developed theory and provides an independent and often
useful route for solving dynamical equations.$\rm{^1}$  In
particular, for periodic motion, the action variable enables
one  to  obtain  the frequencies of a given system directly
without having  to  solve  the equations of motion completely.
The quantum H-J theory has also been studied since the inception
of quantum mechanics.$\rm ^{2}$ It has been recently shown that,
analogous to the classical periodic systems, the quantum action
variable can be profitably employed to arrive at the energy
eigenvalues for potential problems, without obtaining the
corresponding wave functions.$\rm ^{3,4}$ This is to be
contrasted  with  the  standard  procedure to tackle bound state
problems, where, the Schr\"odinger equation is solved both for
the eigenvalues and eigenfunctions. \\ 
The  whole  approach  of the  quantum  H-J  theory  to the
potential  problems  is quite elegant, requiring only some
knowledge  of complex  variables.  Keeping in mind that a
student with this background will be able to appreciate it, we
have made this article quite  pedagogical and  self-contained.
In  Sec. II, we briefly outline the quantum H-J formalism and
its connection with the Schr\"odinger equation and work out the
familiar harmonic oscillator example explicitly. Section III is
devoted to the study of some known  and not so well known
potential problems to elucidate the power of this method. In
Sec. IV, we show why approximate quantization schemes such as, WKB
and supersymmetric (SUSY) WKB, give exact answers for certain
potentials.  We end this paper with some discussions and
concluding remarks. For convenience, a table giving 
relevant information about the potential problems has also been
included. \\
\smallskip 
\newpage 
\noindent
{\bf {II. QUANTUM HAMILTON-JACOBI FORMALISM} } \\ 
\smallskip 
Quantum H-J formalism has been developed along the lines of the classical 
H-J theory since the beginning of quantum mechanics, by the pioneers of the 
field.$\rm^{2}$ In fact, this approach  was christened as the 
``Royal road to quantization'' in the early days of quantum 
mechanics.$\rm^{5}$ In  1983,  Leacock  and  Padgett$\rm^{3,4}$         
demonstrated  that  the quantum H-J formalism can yield the {\it exact} 
eigenvalues for the potential problems provided boundary conditions 
are  applied  judiciously. In what follows, after elaborating 
on the connection of this method with the standard text book approach to 
bound state problems, we solve the harmonic oscillator potential, 
as an illustration. \\
In the conventional approach to non-relativistic stationary state problems, 
one solves the Schr\"odinger equation 
\begin {equation} 
\hat {H} \psi = \left ( \frac{\hat {p}^2}{2m} + V(x) \right ) \psi = E \psi   
\, \, \, \, ,
\end {equation}

\noindent 
for the eigenvalues and eigenfunctions. 
In the quantum H-J formalism, the postulated quantum H-J equation, 
\begin {eqnarray} 
\frac{\hbar}{i} \frac{ \partial ^2 W(x,E)}{ \partial x^2} + 
 \left(\frac{ \partial W(x, E)}{ \partial x} \right)^2 =
\frac{\hbar}{i} \frac{ \partial p(x,E)}{ \partial x} + 
 p^2(x,E) &=& 2m(E - V(x))   \nonumber  \\
 & \equiv & p^2_c(x,E) \, \, \, \, ,
\label{3}
\end {eqnarray}
replaces the Schr\"odinger equation as the dynamical equation. 

\noindent
Here, $W(x,E)$ is  the quantum characteristic function and 
\begin{equation}
p(x,E)
= \frac{\partial W(x,E)}{\partial x}\,\,\, ,
\end{equation}

\noindent
is the quantum momentum function (QMF) and $p_c(x,E)$ is defined to
be the classical momentum function:
\begin{equation} 
p(x,E) \stackrel {\hbar \rightarrow 0} {\rightarrow} p_c(x,E) \, \, \, \, .
\label{6} 
\end{equation}  

\noindent
This  can be thought of as a manifestation of the {\it
correspondence principle} or as a {\it boundary condition} on
the QMF. As will be seen explicitly later, the above condition
helps in determining $p(x,E)$ unambiguously. The quantum
characteristic function $W(x,E)$ is related to the energy
eigenvectors in the coordinate representation as,
\begin{equation}
\psi (x,E) = <x|E> = e^{\frac{iW(x,E)}{\hbar}} \,\,\, ,
\end{equation}

\noindent
and in the same representation,
\begin{eqnarray}
<x|\hat p |E> = -i \hbar \frac{\partial}{\partial x} <x|E> 
&=& \frac{\partial W}{\partial x}<x|E>  \, \, \, \,  \nonumber \\ 
&=&  p(x,E) <x|E> \, \, \, \, .
\end{eqnarray}

\noindent
Thus one gets,
\begin{equation}
p(x,E)= \frac {\hbar}{i} \frac {1} {\psi} 
\frac {\partial \psi (x,E)} {\partial x} \, \, \, \, .
\label{8}
\end{equation}

\noindent 
It is straightforward to check that the Schr\"odinger equation 
goes over to the corresponding quantum H-J equation under the above 
substitution and vice versa.\\ 
The quantum analog of the classical action variable is defined as  
\begin{equation} 
 J(E) \equiv (1/2\pi)\oint_Cdx \, p(x,E) \, \, \, \, .
\label{7}
\end{equation} 

\noindent
Here, $C$ is a counter clockwise contour in the complex $x-$plane, enclosing 
the real line  between the classical turning points. The turning points 
between which the classical motion takes place, are the real values of $x$, 
for which $p^2_c(x,E)$ vanishes. The wave function is known to
have nodes between the classical turning points. These nodes
correspond to poles of the quantum momentum function. To see
this clearly, near a zero of the wave function, located at $x_0$,
we write,
\begin{equation}
\psi = (x - x_0) \phi (x)  \, \, \, \, .
\end{equation} 
This implies, 
\begin{equation}
p \approx \frac{\hbar }{i} \, \, \, \frac{1}{x-x_0}  + \cdots \, \, \, \, .
\label{9}                               
\end{equation} 

\noindent
It is thus seen that, $ p(x,E) $ has a first order pole at $x_0$
with residue $- i \hbar  $. One can also verify the correctness of
Eq. ($\ref{9}$) directly from the quantum H-J equation. Substituting
Eq. ($\ref{9}$) in Eq. ($\ref{3}$), one sees that the contributions of
these poles from $p^2(x,E)$ and $ -i \hbar \partial p(x,E) /
\partial x $ cancel each other only if these poles are of first
order, each having the residue $-i\hbar$. The first order 
poles are of quantum mechanical origin and their positions are
energy dependent, being the same as the zeros of the energy
eigenfunction. Just as the zeros of the wave function change
their positions with energy, so do the location of the
corresponding poles in the QMF. These poles will be referred to
as the moving poles.
\noindent 
The quantum H-J equation shows that, $p(x,E)$ can have
singularities in the complex $x-$plane, other than the moving poles
on the real axis. These singular points correspond to  the
singular points of the potential term $V(x)$.  The
locations of these singularities are, quite obviously,
independent of energy; these poles will be called fixed poles. \\  
\noindent 
For a given energy level the quantum number `$n$' equals the
number of nodes of the wave function and hence it counts the
number of moving poles of $p(x,E)$ inside the contour C appearing
in the definition of the quantum action variable given by Eq.
($\ref {7}$). The residue of each of these poles is $- i \hbar$.
Hence, we have
\begin{equation} 
J(E) = n\hbar \, \, \, \, ,
\label{8.1}
\end{equation}
as an exact quantization condition. \\  
\noindent
Even though this approach appears similar to that of the
familiar WKB scheme, it is worth pointing out that,
Eq. ($\ref{8.1}$), when inverted for $E$ reproduces the {\it exact}
quantized energy eigenvalues.
\noindent
Although a priori, the location and the number of the moving poles
are not known, a suitable deformation of the contour in the
complex plane and  change of variables allows one to compute
$J(E)$, for many potentials, in terms of the fixed poles whose
locations and residues are known. In what follows, explicit
examples are given to clarify this method.
\newpage
\noindent
{\bf { Examples:} } \\
{\it {1. Harmonic oscillator}} \\ 
\noindent 
The quantum H-J equation for the harmonic oscillator problem with  
$V(x) = m \omega^2 x^2 /2 $ is, 
\begin{equation}
p^2 + \frac{\hbar}{i} \frac{ \partial p(x,E)}{ \partial x} 
 = 2m(E - \frac{m \omega^2 x^2} {2})\equiv p^2_c \, \, \, .
\end{equation}

\noindent
The turning points, determined from $p^2_c(x,E)=0$, are $-x_1 = x_2 =
+\sqrt{2E /(m\omega^2)}$. The quantization condition is given by
\begin{equation}
 J(E) = (1/2\pi)\oint_C p(x,E) dx = n \hbar \, \, \, \, .
\end{equation}

\noindent
Here $C$ is the contour enclosing the moving poles between the
two turning points $x_1$ and $x_2$ (see Fig. 1).  Noticing that,
there is only one fixed pole of $p(x,E)$ at $x\rightarrow\infty$, 
to evaluate $J(E)$, one considers an integral $I_{\Gamma_R}$
over a circular contour $\Gamma_R$ having radius $R$ and
oriented in the anti-clockwise direction.  The QMF has no
singular points between $\Gamma_R$ and $C$.  Hence, for this
case, $J(E)$ coincides with $I_{\Gamma_R}$:
\begin{equation}
I_{\Gamma_R} = J(E) \, \, \, \, .
\label{3.8c}
\end{equation}

\noindent 
For the evaluation of the contour integral $I_{\Gamma_R}$, 
one makes a change of variable $ x = 1/y $ to get, 
\begin{eqnarray}
I_{\Gamma_R} & =& (1/2\pi)\oint_{\Gamma_R} dx  p(x,E)
\nonumber  \\ 
&=& (1/2\pi)\oint_{\gamma_0} dy \tilde p(y,E)/y^2   \, \, \,  .
\label{3.8b}
\end{eqnarray} 

\noindent
Here, $\tilde p (y,E) = p(1 / y,E)$  and the counter clockwise contour
$\gamma_0$ encloses only one singular point in the $y$ plane,
i.e., the pole at $y=0$. The corresponding contour integral 
can be straightforwardly calculated. Note 
that there is no negative sign before the integral;  the
direction of the contour changes sense under this mapping, which
is compensated by the negative sign coming from the integration
measure. In this example $J(E)$ and $I_{\Gamma_R}$ are equal, 
though the relation between $J(E)$ and
$I_{\Gamma_R}$ will change from one example to another, the 
method of computing $I_{\Gamma_R}$ remains the same for all 
examples.    \\ 
The quantum H-J equation written in the $y$ variable becomes
\begin{equation}
\tilde {p}^2(y,E) + i \hbar y^2 \frac{ \partial \tilde {p}(y,E) }
{ \partial y} = 2m(E - \frac{m \omega^2} {2y^2}) = \tilde
{p}^2_c \, \, \, . 
\label{3.9}
\end{equation}
To calculate the contribution of the pole at $y=0$, $\tilde {p}(y,E)$ 
is expanded in a Laurent series as, 
\begin{equation}
\tilde{p}(y,E) = \sum_{n =0}^{\infty}a_n y^n + 
\sum_{q = 1}^{k}\frac{b_q}{y^q} \, \, \, \, .
\label{3.11}
\end{equation}

\noindent
Making use of the above expansion of $\tilde {p}(y,E)$ in Eq. ($\ref{3.8b}$),  
one notices that, the only non-vanishing contribution comes from the 
coefficient $a_1$ of the linear term in $y$.  \\ 
In the next step, substituting $\tilde {p} (y,E)$ in the quantum H-J equation 
and comparing the left and right hand sides, it is found that, $b_q = 0$
for $q>1$. On equating the coefficients of the different powers of 
$y$, we have, 
\begin{eqnarray}
b_1^2 &=& -m^2\omega^2   \, \, \, \,  , \label{3.12} \\
2a_0b_1 &=& 0  \, \, \, \,   ,    \label {3.13} \\
-i \hbar b_1 + 2a_1b_1 + a^2_0 &=& 2mE  \, \, \, \, . \label{3.14}
\end{eqnarray}
\noindent 
From Eq. ($\ref {3.12}$), one finds $b_1 = \pm im \omega $. This
ambiguity in sign for $b_1$ can be removed, if we apply the
boundary condition given by Eq. ($\ref {6}$). In the convention
followed here, the  classical momentum function is
defined such that $p_c(x,E) = +i|p_c|$ on the positive real
axis$\rm .^4$ In the limit $y \rightarrow 0, (x \rightarrow
\infty)  $, $p_c \approx {im \omega} / {y}$ and therefore from 
Eq. ($\ref{3.11}$) it follows that  $b_1 = im \omega$. From Eq. ($\ref
{3.13}$), we then have $a_0 = 0$. Substituting the value of $b_1$ in
Eq. ($\ref{3.14}$) one gets $a_1= (2E-\hbar
\omega) / (2 i \omega) $. Plugging Eq. ($\ref {3.11}$) in
Eq. ($\ref {3.8b}$) and noting Eq. ($\ref {3.8c}$) gives,  
 
\begin{equation}
J(E) = I_{\Gamma_R} = i a_1 = \frac{2E-\hbar\omega}{2 \omega}  
\, \, \, \, . 
\label{3.15a}
\end{equation}
\noindent
Thus the quantization condition $J(E) = n \hbar $ when inverted
for $E$, gives 
\begin{equation}
E = \left(n + \frac{1}{2}\right)\hbar \omega \, \, \, .
\end{equation}

\noindent 
2. {\it {Harmonic oscillator on half line}}

\noindent
The quantum H-J equation for this case is obviously the same as
that of the previous example, apart from the condition that the
potential is $\infty$ at $x=0$. This forces us to assume a fixed
pole for $p(x,E)$ in  the complex $x$ plane at $x = 0$ and this
also serves as one of the turning points. The other turning
point is located at $x_2=\sqrt {2E /(m \omega^2)}$. We have,
\begin{equation}
J(E) =
(1/2\pi)\oint_C p(x) dx \, \, \, \, \, , 
\end{equation}
where $C$ is the contour enclosing the moving poles
between zero and $\sqrt{ 2E /( m \omega^2)}$. To evaluate $J(E)$, we
consider a contour  integral $I_{\Gamma_R}$ over a circle of
radius $R$ in the complex plane; $R$ is taken large enough to enclose all
the singular points of the QMF inside it.  This contour integral, 
\begin{equation}
I_{\Gamma_R} \equiv (1/2\pi)\oint_{\Gamma_R} p(x) dx  \, \, \, \, \, , 
\end{equation}
gets contributions from the singular points of $p(x,E)$ inside
$\Gamma_R$. These are, (i) fixed pole at $x=0$, (ii) moving
poles between $0$ and $x_2 = \sqrt {2E /( m \omega^2)}$ and
(iii) moving poles on the negative real axis between $0$ and
$x_1 =- \sqrt {2E /( m\omega^2)}$. The poles on the negative
real axis arise due to the symmetry $x\rightarrow -x$ of the
problem. Hence,
\begin{equation}
I_{\Gamma_R} = J(E) + I_{C_1} + I_{\gamma_1}   \, \, \, , 
\label{4.13a}
\end{equation} 

\noindent
where $I_{\gamma_1}$ is the contour integral for the contour
$\gamma_1$ that takes into account the additional pole at $x =
0$ and $I_{\Gamma_R}$ is the contour integral for the contour
$\Gamma_R$ enclosing all the moving and fixed poles of $p(x,E)$.
$I_{C_1}$  is the contour integral for the counter clockwise
contour $C_1$, enclosing the moving poles on the negative side
of the real axis (see Fig. 2).  However, under $x\rightarrow -x$,
the turning points $x_1$ and $x_2$ and the moving poles get
interchanged. Thus,
\begin{equation}  
 I_{C_1} = J(E) \, \, \, \, .
\label{4.11a}
\end{equation} 
\noindent
Substituting the above in Eq. ($\ref {4.13a}$) we have,
\begin{equation}
I_{\Gamma_R} = 2J(E) + I_{\gamma_1}   \, \, \, .
\label{4.13b}
\end{equation} 
\noindent 
Observing that, it is only the $ b_1/x$ term in the Laurent
expansion of $p(x,E)$ that contributes to the contour integral
$I_{\gamma_1}$, the relevant relation involving $b_1$ following
from the quantum H-J equation is,
\begin{equation}
b_1^2 - \frac{\hbar}{i}b_1 = 0 \, \, \, .
\end{equation}

\noindent
From the two solutions $b_1 = 0 ; -i\hbar$, the former is discarded, 
since it does not give rise to the required singularity at $x=0$.   
Hence, using the second value, 
\begin{equation}
I_{\gamma_1} = \hbar \, \, \, \, . 
\label{4.15a}
\end{equation}

\noindent 
$I_{\Gamma_R}$ is evaluated as in the previous section, 
and is found to be exactly the same. We now substitute the values of
$I_{\gamma_1}$ and $I_{\Gamma_R}$ from
Eqs. ($\ref {4.15a}$) and ($\ref {3.15a}$) in Eq. ($\ref {4.13b}$).  
The quantization rule
$J(E)= n \hbar $, when inverted for $E$, gives the energy
eigenvalue equation;
\begin{equation}
E = \left((2n + 1) + \frac{1}{2}\right)\hbar \omega \, \, \, \, .
\end{equation}

\noindent 
{\bf III. OTHER SOLVABLE EXAMPLES} \\ 
In this section, the quantum H-J method will be applied to a
set of trigonometric  and hyperbolic potentials$\rm .^6$ These
potentials have attracted considerable attention in the recent
literature due to the fact that they can be solved algebraically
using the techniques of SUSY quantum mechanics$.^{\rm 7}$ As
will be seen explicitly later, SUSY shows an alternate way of
selecting the correct solutions of the quantum H-J equation.
Suitable exponential mappings are used in the analyses of the
trigonometric and hyperbolic potentials. Here, the example of
the Eckart potential will be worked out in detail and the
relevant information about the rest of the potentials will be
given in Table I. We also work out the infinite square well
potential in this section which has been deliberately tackled at
the end because of its nontrivial nature in context of the
present formalism.

\noindent
{\it 1. Eckart potential} \\ 
The Eckart potential, by suitably adjusting the ground state energy, 
can be written in the form, 
$$
V(x) =  \omega^2(x) - \frac{\hbar}{\sqrt{2m}} 
\frac{\partial \omega(x)}{\partial x} \, \, \, \, , $$

\noindent                
where  
\begin{equation}
\omega (x) = - A \coth \alpha x + B/A \, \,\, \, ,
\label{1.10}
\end{equation}

\noindent 
is known as the superpotential in the literature$.\rm ^7$ Given
the ground state wave function $\psi_0$, in SUSY quantum
mechanics,
\begin{equation}
\omega(x) = - \frac{\hbar}{\sqrt{2m}} \frac {1}{\psi_0}
\frac{\partial \psi_0}{ \partial x} \, \, \, \, .
\label{1.10a}
\end{equation}

\noindent
The QMF, $p(x,E)$, becomes equal to $i \sqrt {2m}\omega (x)$ for $E=0$. The
quantum H-J equation for the Eckart potential is given as,
\begin{equation}
p^2(x,E) + \frac{\hbar}{i}\frac{\partial p(x,E)}{\partial x} = 
2m \left ( E - A^2 - \frac {B^2} {A^2} -A(A-\frac {\alpha \hbar}
{\sqrt{2m}}) {\rm cosech}^2 \alpha x + 2 B \coth \alpha x \right ) \, \, \, ,
\label{1.11}
\end{equation}

\noindent

where $x$ lies on the half line. To simplify the analysis, we
use the mapping $y = \exp(\alpha x)$ and the corresponding
equations ($\ref{1.10}$) and ($\ref{1.11}$) become
\begin{equation}
\omega (y) = - A \frac{y^2+1}{y^2-1} +\frac{B}{A} \, \, \, \, ,
\label{1.12}
\end{equation}
\noindent
and
\begin{equation}
\tilde p^2(y,E) 
+ \frac{\hbar \alpha y}{i}\frac{\partial \tilde p(y,E)}{\partial y} 
= 2m\left( E -A^2 - \frac{B^2}{A^2} 
-\frac{4A(A-\frac {\alpha \hbar} {\sqrt{2m}})y^2}{(y^2-1)^2} 
+ \frac{2B(y^2+1)}{(y^2-1)}\right) \, \, \, ,
\label{1.13}
\end{equation}

\noindent 
respectively. Here $\tilde p(y,E) = p(\log y / \alpha, E)$. The right hand
side of Eq. ($\ref{1.13}$) equated to zero has four solutions, out
of which two turning points are in the physical regime. The
quantization condition in the $y$ variable takes the form,
\begin{equation}
J(E) =  \frac{1}{2\pi\alpha}\oint_{C_1} \frac{dy}{y} \tilde p(y,E) 
= n\hbar \, \,\, .
\label{1.14}
\end{equation}
\noindent 
It is clear from Eqs. ($\ref{1.13}$) and ($\ref
{1.14}$),  that the integrand has singularities at $y=0, \,
\,  \pm 1$. \\ 
As before, we shall now consider a circle $\Gamma_R$ of radius
$R$, which is such that all the singular points of the integrand
in Eq. ($\ref {1.14}$) are enclosed. Then 
\begin{equation}
I_{\Gamma_R} = I_{C_1} + I_{C_2} + I_{\gamma_1} + I_{\gamma_2} + I_{\gamma_3} 
\, \, \, \, .
\label{1.17}
\end{equation}
\noindent
where $I_{C_1} \equiv J(E) $ and $I_{C_2}$ are the integrals
along the counter clockwise contours $C_1$ and $C_2$ enclosing
the classical and non-classical turning points respectively.
$I_{\gamma_1}$, $I_{\gamma_2}$ and $I_{\gamma_3}$ are the integrals
along contours $\gamma_1$, $\gamma_2$ and $\gamma_3$ which
encloses the singular points at $y=1, y=-1$ and $y=0$
respectively (see Fig.  3). It may be noticed that, the symmetry
$y \rightarrow - y$ in Eq. ($\ref{1.13}$) interchanges the turning
points in the non-classical region with those in the classical
region. Thus, 
\begin{equation}
I_{C_1} = I_{C_2} \, \, \, \, \, . 
\label{1.16}
\end{equation} 
Therefore, from Eq. ($\ref {1.17}$), we have 
\begin{equation}
2J(E) = I_{\Gamma_R} - I_{\gamma_1} - I_{\gamma_2} - I_{\gamma_3} 
\, \, \, \, .
\label{1.17b}
\end{equation}
To find the contribution for the pole at $y=1$, one expands
\begin{equation}
\tilde p(y,E) = \frac{b_1}{(y-1)} +a_0 + a_1 (y-1) + \cdots \, \, \, \, ,
\label{1.18}
\end{equation}

\noindent 
and substitutes the same into Eq. ($\ref{1.13}$). Comparing the
coefficients of $1/(y-1)^2$ on both sides gives,
\begin{equation}
b_1 = \frac {-i \alpha \hbar \pm i(\alpha \hbar - 2\sqrt{2m}A)}{2} \, \, \, \, 
.
\label{1.19}
\end{equation}

\noindent 
Similarly for the pole at $y=-1$ an  expansion in powers of $(y+1)$ is 
sought in $\tilde p(y,E)$ and we have
\begin{equation}
b_1^\prime = \frac {i \alpha \hbar \pm i(\alpha \hbar -
2\sqrt{2m}A)}{2}  \, \, \, ,
\label{1.20}
\end{equation}

\noindent 
where $b_1^\prime$ is the coefficient of the $1/(y+1)$ term in the above 
expansion. To find the residue of the integrand at the pole at $y=0$, we 
expand $\tilde p(y,E)$ as
\begin{equation}
\tilde p(y,E) = a_0 + a_1 y + \cdots \, \, \, \, ,
\label{1.21}
\end{equation}
and comparing the coefficient of the constant term gives 
\begin{equation}
a_0^2 = 2m(E - A^2 - \frac{B^2}{A^2} - 2B) \, \, \, \, .
\label{1.22}
\end{equation}
For the calculation of $I_{\Gamma_R}$, one more change of variable in the 
form  of $y = 1/z$ is sought,  so that the singularity at 
$y \rightarrow \infty$ is mapped to the one at $z=0$. 
Proceeding as before, the coefficient of the constant term in
the expansion of $\tilde {\tilde p} (z,E)$ in powers of $z$ is, 
\begin{equation}
a_0^{\prime^2} = 2m(E - A^2 - \frac{B^2}{A^2} + 2B)  \, \, \, \, ,
\label{1.23}
\end{equation}
\noindent 
where $\tilde {\tilde p}(z,E) = \tilde p(1/z,E)$. 
To find the spectrum for the Eckart potential, one has to choose
correct signs for the above coefficients, by appropriate
boundary conditions.  The procedure as originally suggested by
Leacock and Padgett is a bit complicated to implement. Instead,
we adopt an alternate method, where a similar expansion of the
superpotential $\omega(x)$ at the given poles is compared with
the $E \rightarrow 0$  limit of the above coefficients, there by
fixing the correct sign. This is because Eq. ($\ref {1.10a}$)
implies that $p(x,E) =i\sqrt{2m}\omega(x) $ at zero energy.
Therefore from Eqs. ($\ref {1.19}$), ($\ref {1.20}$), 
($\ref {1.22}$) and ($\ref {1.23}$), we have, 
\begin{eqnarray} 
\alpha \, I_{\gamma_1} &=& \sqrt{2m} A  \, \, \, \, , \\ 
\alpha \, I_{\gamma_2} &=& \sqrt{2m} A  \, \, \, \, , \\ 
\alpha \, I_{\gamma_3} &=& -i\sqrt{2m(E- A^2 -\frac{B^2}{A^2} -2B)}  
 \, \, \, \, , \\ 
\alpha \, I_{\Gamma_R} &=& \sqrt{2m(E- A^2 -\frac{B^2}{A^2} +2B)} 
 \, \, \, \, . 
\end{eqnarray}
Taking into account the contribution of the two identical
contours enclosing the moving poles in the complex $y-$plane, we
find 
\begin{equation}
I_{\Gamma_R} - \sum_{p =1}^3 I_{\gamma_p} = 2 n \hbar \, \, \, \, .
\label{1.24}
\end{equation}
Solving for $E$:
\begin{equation}
E_n = A^2 + \frac{B^2}{A^2} - \frac{B^2}{(\frac{n \alpha \hbar}
{\sqrt{2m}} + A)^2} -  
(\frac {n \alpha \hbar} {\sqrt{2m}} + A)^2 \, \, \, \, .
\label{1.25}
\end{equation}

\noindent 
The calculation of eigenvalues for other SUSY potentials proceeds along 
the same lines and the results are summarised in Table I. \\
\noindent 
{\it 2. The square well potential} \\ 
Square well potential of width $L$ is the simplest example of
one-dimensional motion where a particle of mass $m$ experiences
a potential, 
\begin{eqnarray}
V(x) &=& 0 \, \, \, \, \, \, \, \, {\rm for} \, \, \,  0<x<L \,
\, \, \, \, , \nonumber \\   
&=& \infty \, \, \, \, \, \, \, \, \, {\rm for} \, \, \,  x\le 0 
\, \, \, {\rm and } \, \, \, x\ge L  \, \, \, \, \, .
\end{eqnarray} 
The nodes  of the eigenfunctions and hence the moving poles of
$p(x,E)$ are located between $x=0$ and $L$. Let $C$ be a
rectangular contour enclosing all the moving poles (see Fig.4a).
Then the quantization condition is given by
\begin{equation}
J(E) = \frac {1}{2\pi}\oint_C p(x,E) dx = n \hbar \, \, \, \, , 
\end{equation}  
where the QMF obeys the quantum H-J equation, 
\begin{equation}
p^2(x,E) - i \hbar \frac{\partial p(x,E)}{\partial x} = 2mE  \,
\, \, \, . 
\end{equation}
\noindent 
We now use a mapping $z=\exp((2\pi ix)/L)$; the contour $C$ in
the $x-$plane (Fig. 4a) is mapped into the contour $PAQRBSP$ in the
$z-$plane . The moving poles get mapped on the middle arc of
unit radius (Fig. 4b). The quantization condition is now given as \\
\begin{equation}
J(E) = \frac{L}{4 i \pi^2}\oint_{C_1} \frac {\tilde p(z,E) dz}{z}  
= n \hbar\, \, \, \, ,
\label{5.2}         
\end{equation}
\noindent 
where $C_1$ is the contour $PAQRBSP$ of Fig. 4b. 
The quantum H-J equation written in terms of the new variable $z$
is, 
\begin{equation}
\tilde p^2(z.E) + \frac {2 \pi \hbar z}{L} \frac{\partial 
\tilde p(z,E)}{\partial z} = 2 m E  \, \, \, \, , 
\label{5.3}         
\end{equation}
where $\tilde p(z,E) = p(L \log (z)/(2\pi i),E)$. The boundary condition
that, the wave function vanishes at $x=0$ and $x=L$ gives rise
to a pole in $\tilde p(z,E)$ at $z=1$. Let $\gamma$ and $\Gamma$
be the inner and outer {\it full} circles of radii $OA$ and $OB$
respectively, both taken in the anti-clockwise direction. The
integral in Eq. ($\ref {5.2}$) can be written in terms of integrals
over $\Gamma$, $\gamma$ and the contour $P^\prime S^\prime
R^\prime Q^\prime P^\prime$ enclosing the pole at $z=1$. Thus, we
get

\begin{equation} 
J(E) = \frac{L}{4 i \pi^2} \left ( \oint_{\Gamma} 
\frac {\tilde p(z,E) dz}{z} 
- \oint_{\gamma} 
\frac {\tilde p(z,E) dz}{z} 
- \oint_{P^\prime S^\prime R^\prime Q^\prime P^\prime} 
\frac {\tilde p(z,E) dz}{z} \right ) \, \, \, \, .
\label{xyz}
\end{equation}

\noindent 
The first integral is computed by changing variables from $z$ to
$y=1/z$, as was done earlier for $\Gamma_R$. The last two
integrals in the above expressions are calculated as usual by
doing the Laurent expansion of $\tilde p(z,E)$ in powers of $z$
and $z-1$ respectively and we have,
\begin{eqnarray}
 \oint_{\Gamma} 
\frac {\tilde p(z,E) dz}{z} & =& -2 \pi i \sqrt{2mE} \\ 
 \oint_{\gamma} 
\frac {\tilde p(z,E) dz}{z} & =& 2 \pi i \sqrt{2mE} \\ 
 \oint_{P^\prime S^\prime R^\prime Q^\prime P^\prime} 
\frac {\tilde p(z,E) dz}{z}&=& \frac{4 \pi^2 i \hbar } {L}
\, \, \, \, .
\end{eqnarray}
\noindent 
Substituting the above in Eq. ($\ref {xyz}$), and solving for $E$ we
get, 
\begin{equation}
E = \left( \frac{\pi^2 \hbar^2}{2 m L^2} \right) (n+1)^2 \, \,
\, \, , \, \, \, \, \, n = 0,1,2, \cdots \, \, \, \, \, .
\end{equation}

\noindent 
{\bf IV. EXACTNESS OF ORDINARY AND SUSY WKB APPROXIMATION SCHEMES} \\  

\noindent 
Semiclassical quantization schemes like WKB have been 
very useful since the early days of quantum mechanics, in getting the 
approximate spectra of bound state problems. Interestingly, for certain 
potentials, these approximation schemes give exact answers. It is well known 
that WKB quantization condition,
\begin{equation}
\int_{x_1}^{x_2} \sqrt{2m(E-V(x))} dx = (n +\frac{1}{2}) \pi \hbar 
\, \, \, \, ,
\end{equation}

\noindent 
reproduces the exact eigenvalues for the harmonic oscillator potential. 
Here, $x_1$ and $x_2$ are the two classical turning points with
$x_1 < x_2$ and $n$ takes 
positive integral values. \\
With the advent of SUSY quantum mechanics it was found that, for
a potential $V(x) = \omega^2(x) -
\frac {\hbar} {\sqrt{2m}}  \frac{\partial \omega (x)}{\partial
x}$, for which the ground state energy is zero; a variation of
the WKB approximation,
\begin{equation}
\int_{x_1}^{x_2} \sqrt{2m(E-\omega^2(x))} = n \hbar \pi; \hspace{1.5in} 
n=0,1,2,... \, \, \, ,\label{4.5a}
\end{equation}

\noindent 
gives exact eigenvalues for a large class of potentials$\rm
.^{8-10}$ Here $x_1$ and $x_2$ are solutions of $E-\omega^2 = 0$
with $x_1<x_2$. This is the well known  SUSY WKB approximation
scheme. Since quantum H-J formalism  gives exact results, and is
very similar to these schemes, it can be used to gain an
understanding as to why  SUSY WKB approximation reproduces the
exact answers for these potentials. \\ 
\noindent
For the harmonic oscillator, where WKB is exact, it can be shown
that WKB and SUSY WKB approximations are identical. Therefore, we
concentrate only on the SUSY WKB scheme here. \, \, In order to
compare SUSY WKB with the quantum H-J quantization condition,
the line integral will be converted to a suitable contour
integration over the complex $x-$plane, with a counter clockwise
contour $C$, enclosing the turning points $x_1$ and $x_2$ 
$\rm .^{11}$ Writing,
\begin{equation}
\frac{1}{\pi}\int_{x_1}^{x_2} \sqrt{2m(E-\omega^2)}dx = 
\frac{1}{2\pi}\oint_C \sqrt{2m(E-\omega^2)}dx= n \hbar \, \, \, \, , 
\label{1.9}
\end{equation}
\noindent 
the contour integral in Eq. ($\ref {1.9}$) can be evaluated, for all the
potentials tabulated in Table I. Amazingly, for all these
cases, the singularity structure of $\sqrt{2m(E-\omega^2)}$, other
than the branch cuts, matches exactly with that of the fixed
poles of $p(x,E)$ in the quantum H-J formalism; the
location of the fixed poles and the corresponding residues are
identical for both the cases$\rm .^{12}$ As has been seen earlier, these
poles and their residues determine the eigenspectra completely.
Hence, it is not surprising that SUSY WKB gives exact answers for
these potentials.  \\ 
\noindent 
     {\bf V. CONCLUSION: }  \\
\noindent 
We have explicitly worked out the eigenvalues  for several
solvable potentials in one dimension, using the quantum H-J
method.  Table I contains all the relevant details about the
steps involved in the calculation. Interestingly, the fixed
poles of the quantum momentum function, which are not of 
quantum mechanical origin, in conjunction with the boundary
conditions, completely determine the spectra for these examples.
The main effort involved  in use  of  this  scheme,  lies in
selecting the correct   roots  for  the residues needed. This
was done, first, by using the boundary condition on the QMF
$p(x,E)$, viz., $p \rightarrow p_c$ in the
limit $\hbar \rightarrow 0$. It was then shown that for the SUSY
potentials, comparison of the answers obtained from the quantum
H-J for $E=0$ with that obtained from the superpotential also
reproduced the correct roots. \\ 
For the potentials, where
certain approximate quantization schemes are exact, this
approach  has provided some useful insight: It is the similarity
of the singularity structure of $p(x,E)$ and $\sqrt{2m(E-\omega^2)}$
in the non-classical regions of the $x-$plane,
namely the matching of the poles and the residues that is
responsible for this exactness.\\ 
Although, we have not dealt
with them here, for the non-exact potentials, it is the
inability to deform the contour appropriately because of the
presence of other poles and branch cuts in the complex $x-$plane
that prevents an exact solution in this approach$\rm .^{12}$
Since, exactly solvable potentials are few and far between, this
method can be potentially useful to construct new ones. \\ 
{\bf Acknowledgements:} Part of this work was completed while two of 
us (RSB and AKK) were visiting The Mehta Research Institute. We
would like to thank Director, Mehta Research Institute,
Allahabad, for support during our stay.  We acknowledge useful
discussions with Profs.  V.  Srinivasan, S.  Chaturvedi, Dr. C.
Nagraj Kumar and N. Gurappa. \\

\noindent{\bf References}
\begin{enumerate}
\item H. Goldstein, {\it Classical Mechanics} (Addison-Wesley, New York, 
1980).
\item P.A.M. Dirac, {\it The Principles of Quantum Mechanics}
(Oxford University Press, London, 1958); 
for some original articles and references see, \\ 
J. Schwinger, {\it Quantum Electrodynamics} (Dover, New York, 1958).
\item R.A. Leacock and M.J. Padgett, Phys. Rev. Lett. {\bf 50}, 3 (1983). 
\item R.A. Leacock and M.J. Padgett, Phys. Rev. D {\bf 28}, 2491 (1983). 
\item A. Sommerfeld, {\it Atomic Structure and Spectral Lines}, 
( E.P. Dutton, New York, 1934) translation by H.L. Brose. 
\item R.S. Bhalla, A.K. Kapoor  and P.K. Panigrahi, {\it Energy
Eigenvalues For Supersymmetric Potentials Via Quantum
Hamilton-Jacobi Formalism,} (hep-th/9507154). 
\item R. Dutt, A. Khare, and U.P. Sukhatme, Am. J. Phys. {\bf 56(2)}, 
163 (1988); \\ F. Cooper, A. Khare and U.P. Sukhatme, Phys. Rep.
{\bf 251}, 267 (1995) and references contained therein.  
\item A. Comtet, A. Bandrauk and D.K. Campbell, Phys. Lett. B {\bf 150}, 
159 (1985). 
\item A. Khare, Phys. Lett. B {\bf 161}, 131 (1985).
\item K. Raghunathan, M. Seetharaman and S.S. Vasan, Phys. Lett.
B {\bf 188}, 351 (1987). 
\item See, for example, page 474 of Ref.{\bf 1} for evaluation of these type 
of integrals.
\item R.S. Bhalla, A.K. Kapoor and P.K. Panigrahi, {\it On The Exactness Of
Supersymmetric WKB Approximation Scheme}, (University of Hyderabad preprint,
December, 1995).

\end{enumerate}

\pagebreak

\pagestyle{empty}
\topmargin=-1truein
\textwidth=7.5truein 
\textheight=9truein
\flushleft
\setlength{\oddsidemargin}{-0.25 in}

Table I :  Hyperbolic and trigonometric potentials. \\
The mapping used for hyperbolic potentials is $y=\exp(\alpha x)$, while \\ 
for the trigonometric ones $y=\exp(i \alpha x)$ is being used; $\alpha$ is 
real 
and  positive.\\
\samepage
\begin{tabular}{ l l c l l }
\hline \hline
Name of & Potential & Fixed 
& $\alpha I_{\gamma_p}$ & Eigenvalue \\
potential & &  poles at
&  &  \\
\hline
 & & & & \\
 &  &   0 & $ -i \sqrt{2m(E-A^2)}$ &    \\ 
Scarf II  & $A^2 + (B^2-A^2-\frac{A\alpha \hbar}{\sqrt{2m}})\times $ &  $i$ & 
$ \sqrt {2m} (iB -A)$ &   \\ 
(hyperbolic) & 
$ {\rm sech ^2}\,\alpha x + B(2A + \frac{\alpha \hbar}{\sqrt{2m}} ) \times $
&$-i$ &$-\sqrt {2m}(iB+A)$ & $\! \! \! \! \! \! \! \! A^2 -(A -\frac{n\alpha 
\hbar}{\sqrt{2m}})^2 $   \\
 & $ {\rm sech}\,\alpha x \, \, {\rm tanh} \,\alpha x$
   & $ \infty $   & $ i \sqrt{2m(E-A^2)} $ &     \\ 

 & & &  & \\
 & & &  & \\
Rosen - & $A^2 + \frac{B^2}{A^2}$  & $0$ & $ - i \sqrt{2m(E-A^2 -\frac{B^2}{A^2} +2B)}$ & 
$ A^2 + B^2 / A^2   $ \\    
  Morse II & $ -A(A +\frac{\alpha \hbar}{\sqrt{2m}} ) \, {\rm
sech^2} \,\alpha x $ & $i$ & $-\sqrt{2m}A$ & $ - (A -
\frac{n\alpha \hbar}{\sqrt{2m}} )^2 $ \\ 
(hyperbolic) & $ + 2B {\rm tanh} \,\alpha x$
 &  $-i$   & $ - \sqrt{2m}A $ & $\! \! \! \! \! \! \! - B^2/(A-\frac{n\alpha 
\hbar}{\sqrt{2m}})^2$   \\     
 &  &  $\infty $ & $ i \sqrt{2m(E-A^2 -\frac{B^2}{A^2} - 2B)} $ &   \\ 
 & & & & \\ 
 & & & & \\
Generalised& 
$A^2 + (B^2+A^2+\frac{A\alpha \hbar}{\sqrt{2m}} )\times$
 &  $0$   & $ - i \sqrt{2m(E-A^2)}$ &   \\ 
Poschl-& ${\rm cosech}^2\,\alpha x - B(2A + \frac{\alpha
\hbar}{\sqrt{2m}} ) \times $ 
&$1$& $-\sqrt{2m}(A-B)$ & $\! \! \! \! \! \! \! A^2 -(A -  \frac{n\alpha
\hbar}{\sqrt{2m}} )^2  $ \\ 
Teller & $ {\rm coth} \,\alpha x \, {\rm cosech}\,\alpha x $
   & $-1$   & $ -\sqrt{2m}(A+B) $ &  \\ 
(hyperbolic)&$( x\geq 0)$ & $\infty$ & $ i \sqrt{2m(E-A^2)} $ &\\
 & & & & \\
 & & & & \\
Scarf I  & 
 $ -A^2 + (A^2+B^2+\frac{A\alpha \hbar}{\sqrt{2m}})\times$ & 
  $0$   & $   \sqrt{2m(E+A^2)}$ &  \\  
(trig-& $\sec^2\,\alpha x - B(2A + \frac{\alpha
\hbar}{\sqrt{2m}}) \times $ & $i$  &$-\sqrt{2m}(A-B) $ & 
$\! \! \! \! \! \! \! (A + \frac{n\alpha
\hbar}{\sqrt{2m}} )^2 - A^2 $ \\
onometric) & $ \sec\,\alpha x \, \tan\,\alpha x$ & $-i$   
& $ -\sqrt{2m}(A + B) $ &   
\\ & ($- \pi/2 \le \alpha x \le \pi /2 $)  &$ \infty $ & $
-\sqrt{2m(E+A^2)}$ & \\ 
 & & & & \\
 & & & & \\
Rosen- & $ A(A -\frac{\alpha \hbar}{\sqrt{2m}} ) {\rm cosec}^2\,\alpha x $
  &  $0$   & $ - \sqrt{2m(E+A^2 -\frac{B^2}{A^2} +2iB)}$ &  
$ A^2 + B^2 / A^2  $ \\ 
Morse-I & $- A^2 + \frac{B^2}{A^2}  $ &$1$ &
$\sqrt{2m}A$ & $ \! \! \! \! \! \! \!  -(A +\frac{n\alpha
\hbar}{\sqrt{2m}} )^2 $ \\ (trig- & $ + 2B \cot \,\alpha x$ &
$-1$ & $ \sqrt{2m}A $ & $ \! \! \! \! \! \! \! \! -B^2/(A+
\frac{n\alpha
\hbar}{\sqrt{2m}})^2$  \\   
onometric)& ($0 \le \alpha x \le \pi$) & 
 $\infty $ & $  \sqrt{2m(E+A^2 -\frac{B^2}{A^2} -2iB)} $ & \\
\hline 
\hline 
\end{tabular}

\pagebreak

{\bf Figure Captions} \\ 
\begin{enumerate}
\item  Figure 1. Contour for the harmonic oscillator problem. \\ 
\item  Figure 2. Contour for the harmonic oscillator problem on
half line. \\ 
\item  Figure 3. Contour for the Eckart potential problem, using
the mapping $y=\exp (\alpha x)$. \\ 
\item  Figure 4a. Contour for the square well problem, in the
$x-$plane. \\ 
\item  Figure 4b. Contour for the square well problem, in the
$z-$plane. 

\end{enumerate}

\end{document}